\documentclass[journal,10pt]{IEEEtran}
\ifCLASSINFOpdf
  \usepackage[pdftex]{graphicx}
  \usepackage{amsthm,amsmath,amssymb}
\usepackage{makecell}
\usepackage{mathrsfs}
  \usepackage{float}
  \usepackage{algorithm}
\usepackage{algpseudocode}
\usepackage{amsmath}
\usepackage{bm}
\usepackage{color}
\usepackage{comment}
\usepackage{booktabs}
 
\else
\fi
\usepackage{leftidx}
\usepackage{epstopdf}
\usepackage{amsmath}
\usepackage{subfigure}
\usepackage{makecell}
\usepackage{multirow}
\begin{document}

\title{Scatterer Recognition for Multi-Modal Intelligent Vehicular Channel Modeling via Synesthesia of Machines}

 \author{Ziwei~Huang,~\IEEEmembership{Member,~IEEE}, Lu~Bai,~\IEEEmembership{Member,~IEEE}, Zengrui~Han,~\IEEEmembership{Graduate~Student~Member,~IEEE},   and Xiang~Cheng,~\IEEEmembership{Fellow,~IEEE}

\thanks{Z.~Huang, Z.~Han, and X.~Cheng are with the State Key Laboratory of Advanced Optical Communication Systems and Networks, School of Electronics, Peking University, Beijing, 100871, P. R. China (email: ziweihuang@pku.edu.cn, zengruihan@stu.pku.edu.cn, xiangcheng@pku.edu.cn).}
\thanks{L. Bai is with the Joint SDU-NTU Centre for Artificial Intelligence Research (C-FAIR), Shandong University, Jinan, 250101, P. R. China (e-mail: lubai@sdu.edu.cn).}
}

\markboth{}
{Zeng \MakeLowercase{\textit{et al.}}: Bare Demo of IEEEtran.cls for IEEE Journals}

		\maketitle

\begin{abstract}
In this paper, a novel multi-modal intelligent  vehicular channel model is proposed by scatterer recognition from light detection and ranging (LiDAR) point clouds via Synesthesia of Machines (SoM). The proposed model can support the design of intelligent transportation systems (ITSs). To provide a robust data foundation, a new intelligent sensing-communication integration dataset in vehicular urban scenarios is constructed. Based on the constructed dataset, the complex SoM mechanism, i.e., mapping relationship between scatterers in  electromagnetic space and LiDAR point clouds in  physical environment, is explored via multilayer perceptron (MLP) in consideration of    electromagnetic propagation mechanism. 
By using LiDAR point clouds to implement scatterer recognition, channel non-stationarity and consistency are captured closely coupled with the environment. Using ray-tracing (RT)-based results as the ground truth, the scatterer recognition accuracy exceeds  90\%. The accuracy of the proposed model is further verified by the close fit between simulation results and RT results.
\end{abstract}

\IEEEpeerreviewmaketitle

\section{Introduction}
\IEEEPARstart {T}{o} support precise localization sensing and efficient communication link establishment for smart vehicles in intelligent transportation systems (ITSs), it is essential to achieve an in-depth understanding of the surrounding environment and high-precision vehicular channel modeling \cite{6G1}. However, widely used approaches, which solely utilize the uni-modal radio frequency (RF) communication information, are difficult to achieve high-precision vehicular channel modeling, and thus cannot support the aforementioned application related to smart vehicles in ITSs.
Fortunately, smart vehicles are equipped with multi-modal devices, which can acquire surrounding environmental information and further assist in  vehicular channel modeling  \cite{LA-GBSM}. To adequately utilize the multi-modal information in the surrounding environment, inspired by human synesthesia,  a novel concept, i.e., Synesthesia of Machines (SoM), is proposed \cite{som}. SoM aims to achieve intelligent integration of communications and sensing via artificial neural networks. As the cornerstone of SoM research, the exploration of SoM mechanism, i.e., mapping relationship between physical environment and electromagnetic space, is essential. Based on the SoM mechanism, a high-precision multi-modal intelligent vehicular channel model can be constructed  by intelligently processing the multi-modal information from communication devices and various sensors.


\begin{figure*}[!t]
	\centering
\includegraphics[width=0.99\textwidth]{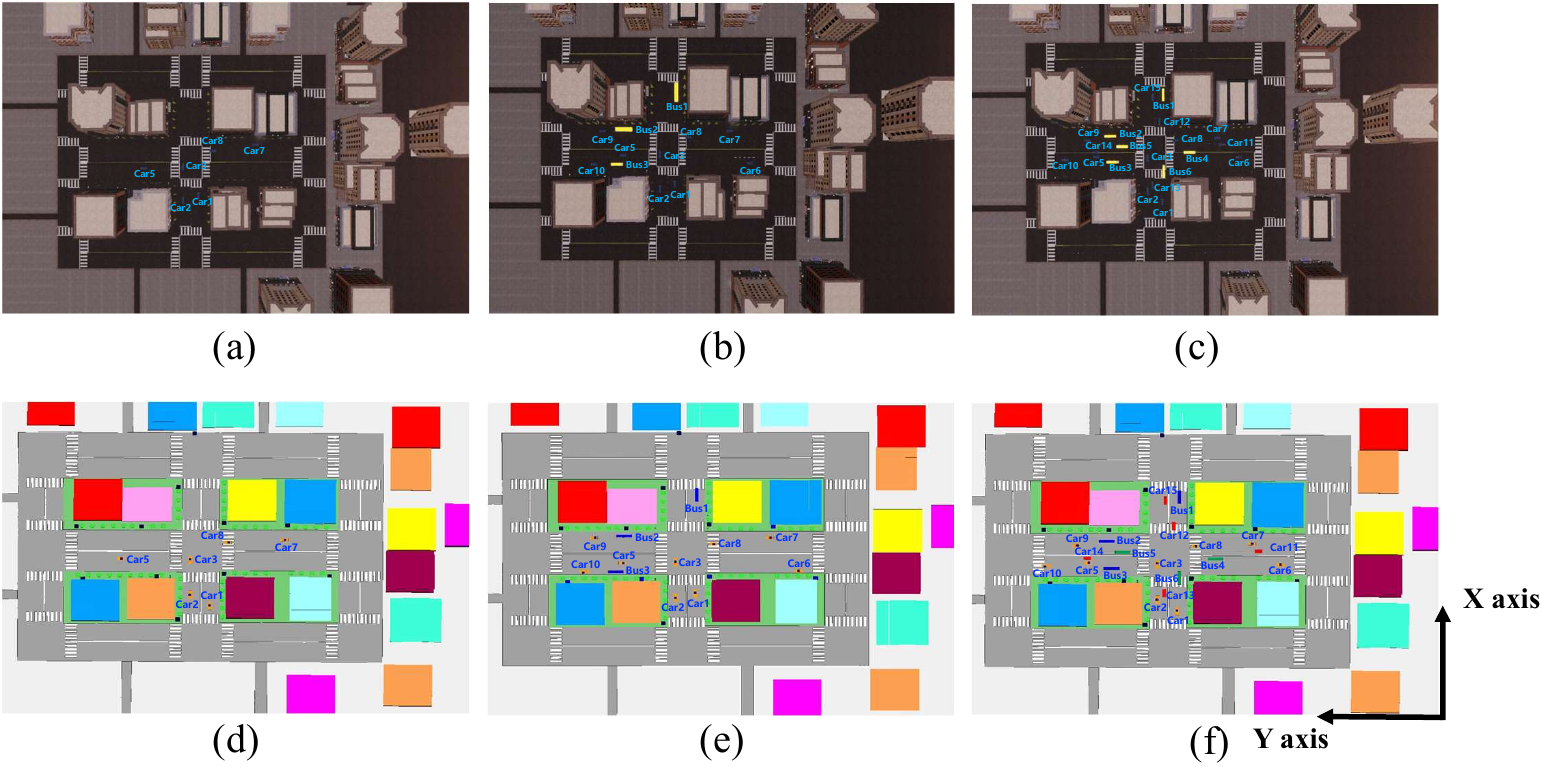}
	\caption{BEV of vehicular urban crossroad simulation scenarios at Snapshot 200. Figs. (a)–(c) are scenarios in AirSim with low, medium, and high VTDs, respectively. Figs. (d)–(f) are scenarios in Wireless InSite with low, medium, and high VTDs, respectively.}
	\label{scenario}
\end{figure*}
Considering the necessity of exploring SoM mechanism, i.e., mapping relationship, some preliminary work has been conducted. The authors in \cite{mapping1} proposed an  environment reconstruction method and explored the mapping relationship between light detection and ranging (LiDAR)  point clouds and path loss. The authors in \cite{mapping2} further explored the mapping relationship between the satellite image and path loss, where environmental features were extracted. However, the  mapping relationships explored in \cite{mapping1,mapping2} were limited to  sensing  and channel large-scale fading. Multipath fading, i.e., channel small-scale fading,  is a significant factor, which affects communication system design and presents more challenges compared to channel large-scale fading \cite{survey1}. To intuitively characterize multipath fading, the concept of scatterers is introduced to model the interaction between radio waves and objects \cite{xilaren}--\cite{COST 2100}. Currently, extensive vehicular channel measurements \cite{mea1}--\cite{mea3} and standardized channel models \cite{COST 2100, 3GPP} have been conducted to explore spatial attributes of scatterers, including their numbers and positions. By characterizing the spatial attributes of scatterers,  channel non-stationarity and consistency can be modeled through birth-death (BD) process and visibility region (VR) \cite{3GPP}--\cite{model2}. Based on the Markov chain, the BD process characterizes the \emph{mathematical relationship} for the variation of the scatterer number, thus capturing channel non-stationarity. Based on the geometry, the VR characterizes the \emph{spatial relationship} for the smooth evolution of scatterers, thus capturing channel non-stationarity and consistency. Nevertheless, the aforementioned two methods focus on modeling the scatterer variation/evolution statistically. Since the \emph{mapping relationship} between objects in  physical environment and scatterers in  electromagnetic space is ignored, channel non-stationarity and consistency cannot be accurately captured. This results in the inability to model the tight interplay between physical environment and electromagnetic space. Although vehicular channel models \cite{3GPP}--\cite{model2} captured the variation/evolution of scatterers and channel non-stationarity/consistency by BD process and VR method, they cannot meet the high-precision requirement due to the consideration of uni-modal  RF communication information. Our previous work in \cite{LA-GBSM} attempted to utilize the multi-modal information from communication and LiDAR devices, and further proposed a LiDAR-aided geometry-based stochastic modeling (LA-GBSM). However, the mapping relationship between physical environment and electromagnetic space was ignored, and thus the capturing of channel non-stationarity and consistency possessed limited coupling with the environment \cite{LA-GBSM}. Therefore, it is essential to propose a multi-modal intelligent vehicular channel model, which captures channel non-stationarity and consistency closely coupled with the environment, i.e., environment-channel non-stationarity and consistency, by exploring the mapping relationship.

To fill this gap and support the design of ITSs, we propose a novel multi-modal intelligent  vehicular channel model via SoM, which intelligently processes the multi-modal information from communication and LiDAR devices. By using AirSim \cite{AirSim} and Wireless InSite \cite{WI}, a new  intelligent sensing-communication integration dataset
in vehicular scenarios with low, medium, and high vehicular traffic densities (VTDs) is constructed. The sensing information, i.e., LiDAR point cloud, is intelligently processed by the density-based spatial clustering of applications with noise (DBSCAN) clustering algorithm to extract  physical environment features, which are aligned with communication information collected from electromagnetic space. By further leveraging a typical artificial neural network, i.e., multilayer perceptron (MLP), with electromagnetic propagation mechanisms, the complex SoM mechanism, i.e., mapping relationship between LiDAR point clouds in physical environment and scatterers in electromagnetic space, is explored for the first time. 
To model environment-channel non-stationarity and consistency, physical environment features via LiDAR point clouds are  utilized for scatterer recognition, thus modeling spatial attributes, i.e., numbers and positions, of scatterers  closely coupled with
the environment. Using ray-tracing (RT)-based results as the ground truth, simulation results show that the scatterer recognition accuracy exceeds 90\% in each VTD condition. The accuracy of the proposed multi-modal intelligent
vehicular
model is also verified by the close fit between simulation results and RT results.


\section{Mapping Relationship Exploration: Scatterer Recognition from LiDAR Point Clouds}

\subsection{High-Fidelity Dataset Construction}
By using AirSim \cite{AirSim} and Wireless InSite \cite{WI}, we construct a new dataset in the vehicular urban crossroad. To obtain high-fidelity LiDAR point clouds, simulation scenarios in AirSim are constructed via the advanced three-dimensional (3D) modeling software with the superior rendering effect. To collect high-fidelity scatterer information, Wireless InSite exploits RT technology based on geometrical optics and uniform theory of diffraction. Similar to our previous work in \cite{CC}, physical environment  in AirSim and electromagnetic space in Wireless InSite further achieve in-depth integration and precise alignment. In AirSim, the LiDAR equipped on each vehicle has $16$ channels, $10$~Hz scanning frequency, and $240,000$ points per second, where the upward and downward field of view (FoV) are $15^\circ$ and $-25^\circ$, respectively. In Wireless InSite, the communication device equipped on each vehicle is operated at $28$~GHz carrier frequency with $2$ GHz bandwidth, where numbers of antennas at transmitter (Tx) and receiver (Rx) are $L_\mathrm{T}$ = $L_\mathrm{R}$ = 1. Antennas are placed on the roof of the car and bus at heights of $2$~m and $3$~m, respectively. 
 
\begin{figure*}[!t]
	\centering
\includegraphics[width=0.99\textwidth]{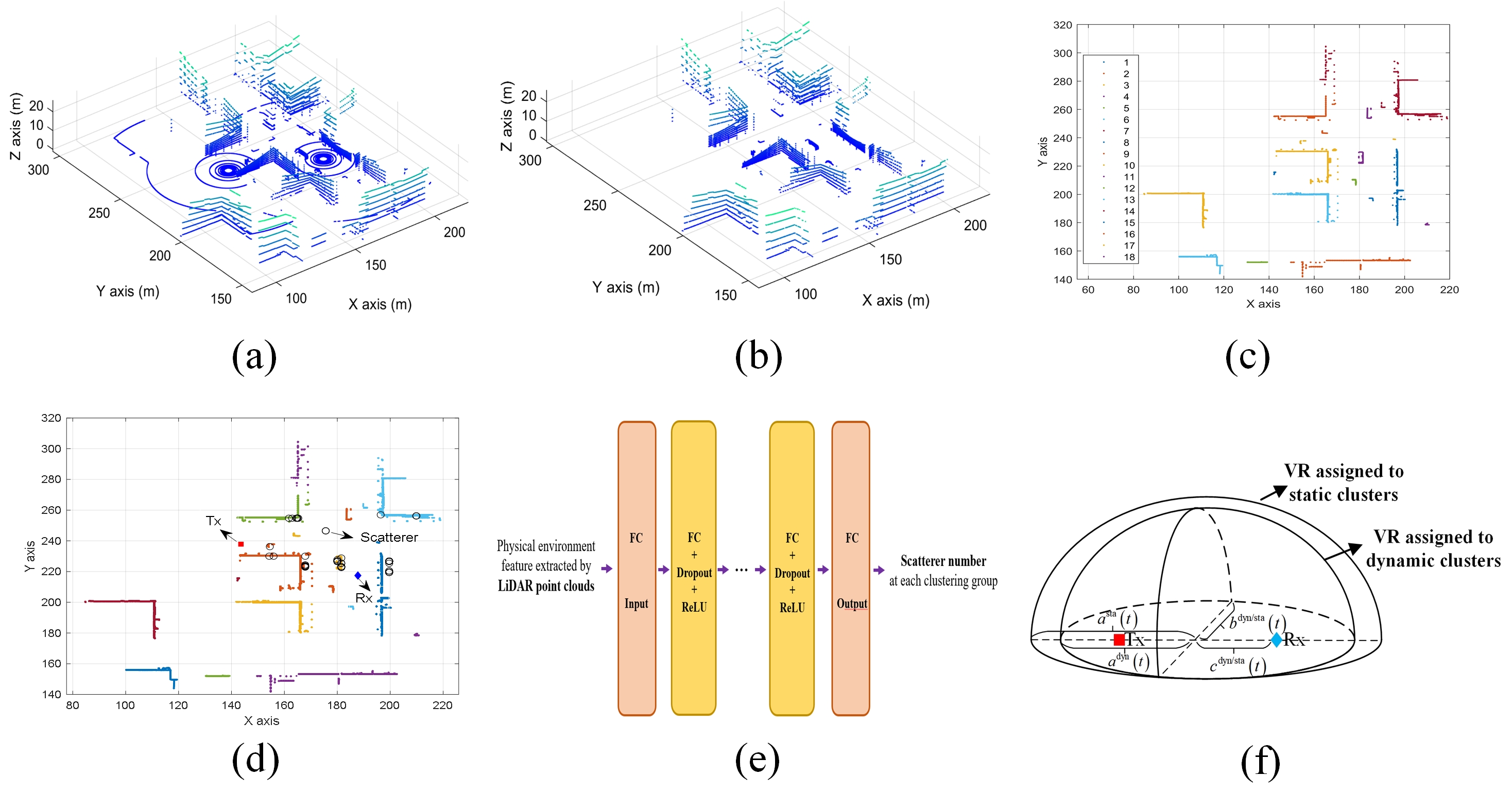}
	\caption{Step list taking Car1 (Tx) and Car8 (Rx) at the crossing street under high VTD at Snapshot 1 as an example.}
	\label{step_list}
\end{figure*}
Given the diversity of the dataset, Fig.~\ref{scenario} demonstrates that we consider three VTD conditions, i.e., low, medium, and high, and three types of streets, i.e., vertical ($x$-axis), horizontal ($y$-axis), and crossing ($xy$-axis) streets. 
Each type of street has $6$ different transceiver links, e.g., Car5 (Tx) and Car7 (Rx) at the horizontal street, Car1 (Tx) and Car2 (Rx) at the vertical street, and Car1 (Tx) and Car8 (Rx) at  the crossing street, thus containing line-of-sight (LoS) and non-LoS (NLoS) conditions. Each transceiver link has $1500$ snapshots and each VTD condition has the same transceiver link. There are $27,000$ snapshots at each VTD condition. Overall, the constructed dataset contains $81,000$ snapshots with high-fidelity LiDAR point clouds and scatterer information.

\subsection{Mapping Relationship Exploration}

For clarity, a step list illustrating the exploration of the SoM mechanism, i.e., mapping relationship between  physical environment and electromagnetic space, is presented below.

\emph{Step 1:} Unlike the monostatic sensing, Tx and Rx have different positions in vehicular communications. Therefore, LiDAR point clouds at Tx and Rx can be concatenated to obtain  physical environment, as shown in  Fig.~\ref{step_list}(a).

\emph{Step 2:} To reduce data redundancy, the ground point is removed by the pre-processing of concatenated LiDAR point clouds, which are further downsampled, as shown in Fig.~\ref{step_list}(b).


\emph{Step 3:} A typical clustering algorithm in machine learning, i.e., DBSCAN, is leveraged to efficiently obtain physical environment features. For clarity, Fig.~\ref{step_list}(c) shows the bird's-eye view (BEV) of LiDAR point clouds, which contain $18$ clustering groups. 


\emph{Step 4:} Since the in-depth integration and precise alignment are conducted in the constructed dataset, physical environment and electromagnetic space can be matched in the same world coordinate system. In Fig.~\ref{step_list}(d), scatterers are  located at the clustering group. According to the RT mechanism, paths are significantly affected by the transmission distance and angle. To calculate the size and orientation of each clustering group, its circumscribed cuboid is obtained. The height of circumscribed cuboid is the same as that of clustering group. The  circumscribed cuboid projection is the minimum perimeter bounding rectangle of the clustering group projection. 
\begin{figure*}[!t]
	\centering
\includegraphics[width=0.99\textwidth]{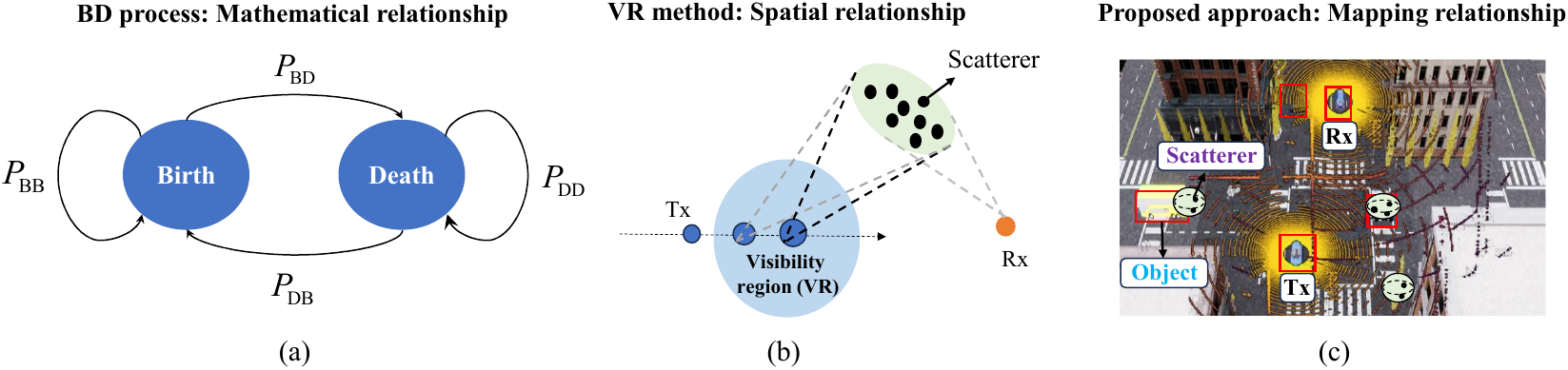}
	\caption{Comparison of the BD process, the VR method, and the proposed approach.}
	\label{Compare}
\end{figure*}

\emph{Step 5:} Considering the advantage of dealing with the task of numerical inputs and numerical outputs,  MLP is exploited to achieve scatterer recognition from LiDAR point clouds, as shown in Fig.~\ref{step_list}(e). The input is physical environment feature extracted by LiDAR point clouds, including the length, width, height, center point, and orientation vector of circumscribed cuboid and the  position of transceiver. The output is scatterer number at each clustering group. For example, in Fig.~\ref{step_list}, the output is a matrix with dimensions of $18$ by $1$. As a result, with the help of MLP, the number of scatterers in electromagnetic space at each  clustering group of LiDAR point clouds in physical environment can be obtained for the first time.  

\begin{figure*}[!t]
	\begin{equation}
		\begin{aligned}
  h(t,\tau)=&\underbrace{\sqrt{\frac{\Omega(t)}{\Omega(t)+1}}
		h^\mathrm{LoS}(t)\delta\left(\tau-\tau^\mathrm{LoS}(t)\right)}_\mathrm{LoS}+\underbrace{\sqrt{\frac{\eta^\mathrm{GR}(t)}{\Omega(t)+1}}h^\mathrm{GR}(t) \delta\left(\tau-\tau^\mathrm{GR}(t)\right)}_\mathrm{Ground \, Reflection}\\
&+\underbrace{\sum_{i=1}^{N_\mathrm{c}(t)}\sum_{n_i=1}^{N_\mathrm{s}(t)}\sqrt{\frac{\eta^\mathrm{sta}(t)}{\Omega(t)+1}}h^\mathrm{sta}_{i,n_i}(t) \delta\left(\tau-\tau^\mathrm{sta}_{i,n_i}(t)\right)+\sum_{j=1}^{M_\mathrm{c}(t)}\sum_{n_j=1}^{M_\mathrm{s}(t)}\sqrt{\frac{\eta^\mathrm{dyn}(t)}{\Omega(t)+1}}h^\mathrm{dyn}_{j,n_j}(t) \delta\left(\tau-\tau^\mathrm{dyn}_{j,n_j}(t)\right)
	}_\mathrm{NLoS}.
\label{desde}
		\end{aligned}
	\end{equation}
 		\hrulefill
\vspace*{4pt}
\end{figure*}
\emph{Step 6:} To further enhance the interpretability of network output, the propagation mechanism is considered via the VR method. Scatterers are divided into dynamic and static scatterers, which are further assigned to VR. Fig.~\ref{step_list}(f) shows the VR assigned to static/dynamic scatterers, i.e., the 3D ellipsoid with the transceiver as the focus, where major axis, minor axis, and focal length are $2a^\mathrm{sta/dyn}(t)$, $2b^\mathrm{sta/dyn}(t)$, and $2c^\mathrm{sta/dyn}(t)$, respectively. VR-related parameters are accurately obtained via RT-based channel data \cite{LA-GBSM}. Finally, scatterers recognized through LiDAR point clouds, which are located outside VR, are deleted and the  output number is also  changed. 

By using the SoM mechanism, i.e., mapping relationship, scatterer recognition from LiDAR point clouds is achieved. This facilitates  intelligent vehicular channel modeling with multi-modal information from communication and LiDAR devices, which achieves the accurate modeling of channel parameters and the precise capturing of environment-channel non-stationarity and consistency.


\section{Multi-Modal Intelligent Vehicular Channel Modeling}

In this section, a high-precision multi-modal intelligent vehicular channel model by scatterer recognition from LiDAR point clouds is proposed to support the design of ITSs. 
The channel impulse response (CIR) is given as \eqref{desde}, where $\Omega(t)$ is Ricean factor. $\eta^\mathrm{GR}(t)$, $\eta^\mathrm{sta}(t)$, and $\eta^\mathrm{dyn}(t)$ are power ratios of ground reflection, static cluster, and dynamic cluster components with $\eta^\mathrm{GR}(t)+\eta^\mathrm{sta}(t)+\eta^\mathrm{dyn}(t)=1$. $h^\mathrm{LoS}(t)$, $h^\mathrm{GR}(t)$, $h^\mathrm{sta}_{i,n_i}(t)$, and $h^\mathrm{dyn}_{j,n_j}(t)$ are complex channel gains of LoS, ground reflection, static cluster, and dynamic cluster components. $\tau^\mathrm{LoS}(t)$, $\tau^\mathrm{GR}(t)$, $\tau^\mathrm{sta}_{i,n_i}(t)$, and $\tau^\mathrm{dyn}_{j,n_j}(t)$ are delays of LoS, ground reflection, static cluster, and dynamic cluster components. $N_\mathrm{c}(t)$/$M_\mathrm{c}(t)$ and $N_\mathrm{s}(t)$/$M_\mathrm{s}(t)$ are numbers of static/dynamic clusters and numbers of rays within static/dynamic clusters, respectively. The detailed generation process of channel parameters in \eqref{desde} is similar to our previous work in  \cite{LA-GBSM}.

Channel non-stationarity and consistency are the typical channel characteristic and feature, which can be captured via BD process and VR method \cite{3GPP}--\cite{model2}. However, since the BD process and VR method model the mathematical relationship and spatial relationship for the scatterer variation, respectively,  the tight interplay between physical environment and channel non-stationarity/consistency cannot be captured. To overcome this limitation and support applications related to smart vehicles in ITSs, the proposed approach exploits multi-modal information from communication and LiDAR devices to achieve scatterer recognition from LiDAR point clouds based on the explored mapping relationship. As a consequence,  environment-channel non-stationarity and consistency are captured. 
For clarity, Fig.~\ref{Compare} illustrates the difference between the BD process, the VR method, and the proposed approach. For the proposed approach, scatterers recognized by LiDAR point clouds essentially correspond to the certain object, such as vehicle, tree, and building, in the proposed approach, which is  different from the BD process and the VR method.

In the proposed approach, at the initial time, the scatterer recognition is implemented by LiDAR point clouds based on the mapping relationship. Similar to \cite{LA-GBSM}, scatterers are divided into static and dynamic scatterers, which are further clustered into static and dynamic clusters. Unlike  \cite{LA-GBSM}, scatterers recognized by LiDAR point clouds correspond to certain objects in physical environment. This leads to accurate channel parameters, including number $N_\mathrm{s}$/$N_\mathrm{c}$/$M_\mathrm{s}$/$M_\mathrm{c}$,  delay $\tau^\mathrm{sta}_{i,n_i}$/$\tau^\mathrm{dyn}_{j,n_j}$, and angle $\alpha^\mathrm{sta}_{i,n_i}$/$\beta^\mathrm{sta}_{i,n_i}$/$\alpha^\mathrm{dyn}_{j,n_j}$/$\beta^\mathrm{dyn}_{j,n_j}$, thus facilitating  multi-modal intelligent vehicular channel modeling.

\begin{figure}[!t]
	\centering
\includegraphics[width=0.49\textwidth]{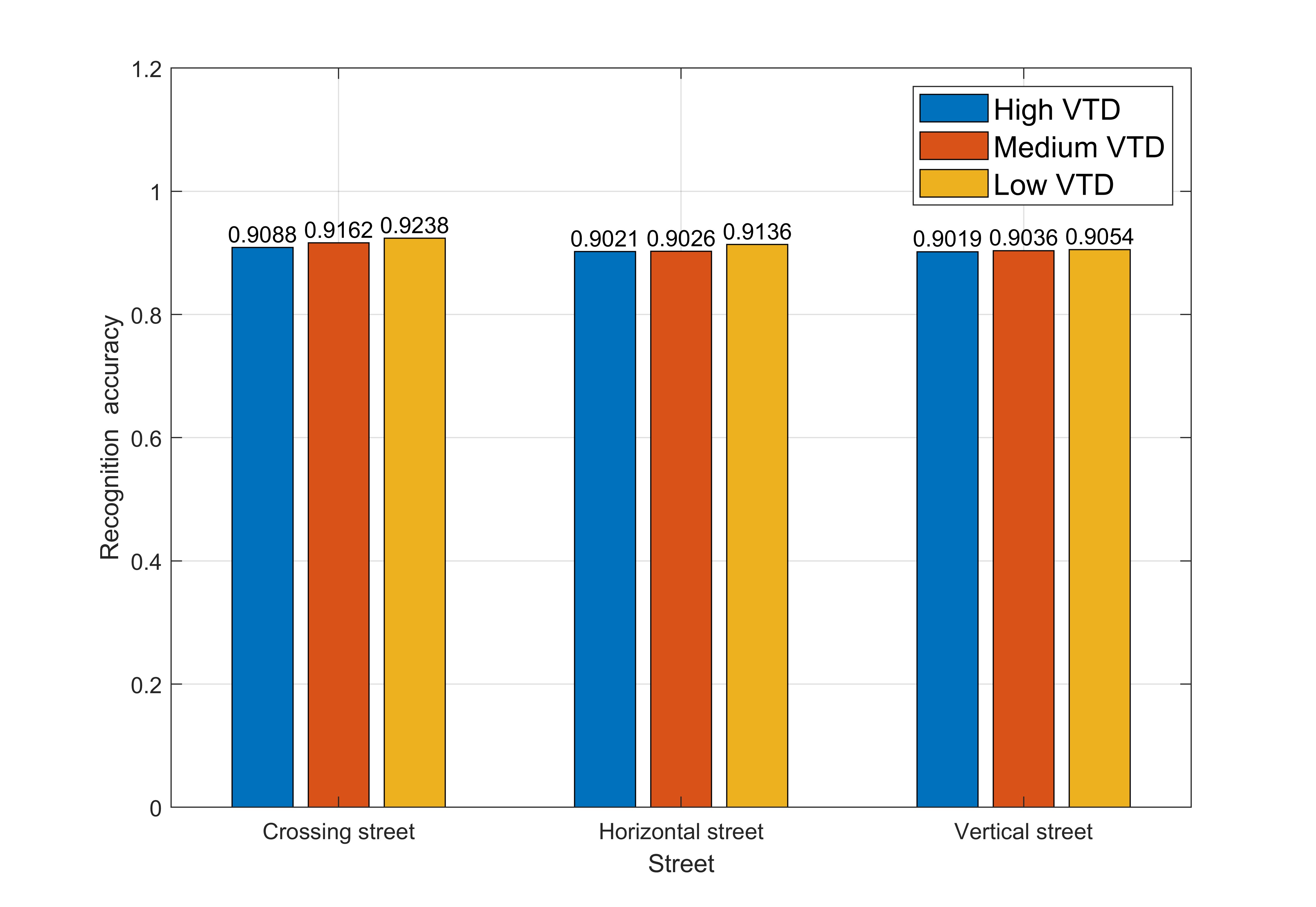}
	\caption{Scatterer recognition accuracy under different VTDs and streets.}
	\label{VTD-street}
\end{figure}
As time evolves and physical environment changes, there are different LiDAR point clouds at different time instants. Through the scatterer recognition from LiDAR point clouds and the capturing of tight interplay between physical environment and  electromagnetic space, scatterers change  with LiDAR point clouds. As a result, environment-channel non-stationarity in the time domain is mimicked. Furthermore, LiDAR point clouds in physical environment at adjacent time instants are similar. In this case, recognized scatterers from LiDAR point clouds are also similar at adjacent time instants, thus capturing  environment-channel consistency in the time domain.
To further model environment-channel non-stationarity and consistency in the frequency domain, a frequency-dependent factor $\left(\frac{f}{f_{c}}\right)^{\chi}$ is introduced to the time-varying transfer function (TVTF). The TVTF is obtained by utilizing the Fourier transform to CIR, which is derived based on the scatterer recognition from LiDAR point clouds with accurate number and position parameters, in respect of delay.  

By intelligently processing multi-modal information from communication and LiDAR devices, the scatterer recognition from LiDAR point clouds is achieved based on the explored mapping relationship. In this case, environment-channel non-stationarity and consistency can be captured, thus achieving high-precision multi-modal intelligent vehicular channel modeling.

\section{Simulation Results and Analysis}
Detailed equipment parameters, e.g., scanning frequency, FoV, carrier frequency, and bandwidth,  for LiDAR point clouds and scatterer acquisition are given in Section II-A. In neural network training, the hyper-parameter setting is listed in Table~\ref{setup2}. The dataset is divided into the training set, validation set, and test set, in the proportion of $3:1:1$. In Figs.~4--6, the accuracy, error probability heat map, and number of scatterer recognition are given to demonstrate high-precision scatterer recognition. The scatterer recognition accuracy of the proposed approach is further compared with that of the existing random generation approach in Fig.~7. To validate the accuracy of the proposed multi-modal intelligent vehicular model, simulation results and RT-based results are compared in Fig.~8. Note that, as the RT technology generates channel data according to the geometrical optical and uniform theory of diffraction, RT-based results are of high precision and are leveraged to validate the proposed model.

\begin{figure}[!t]
	\centering
\includegraphics[width=0.49\textwidth]{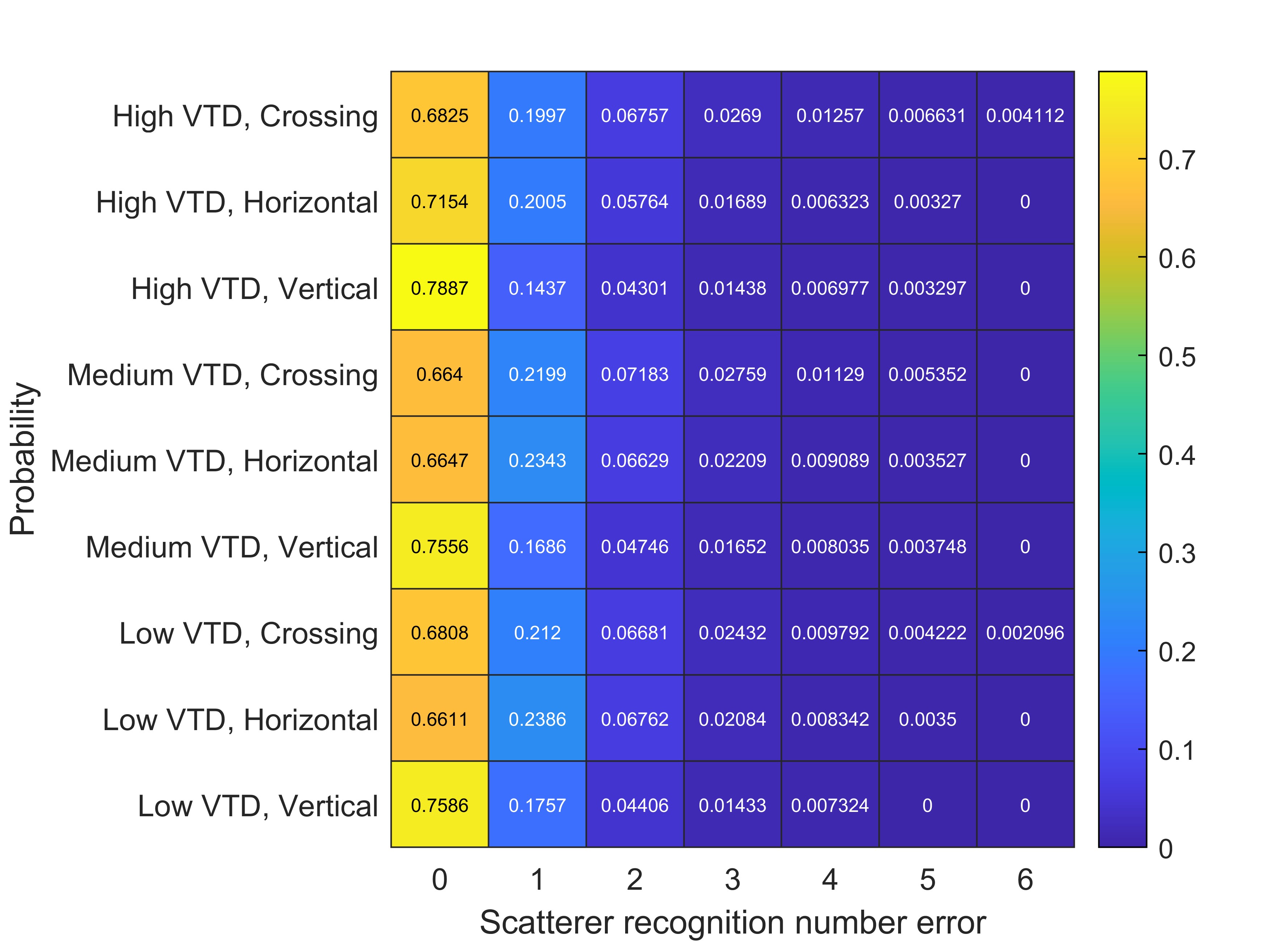}
	\caption{Probability heat map of scatterer recognition number error.}
	\label{Heatmap}
\end{figure}

\begin{figure}[!t]
	\centering
\includegraphics[width=0.49\textwidth]{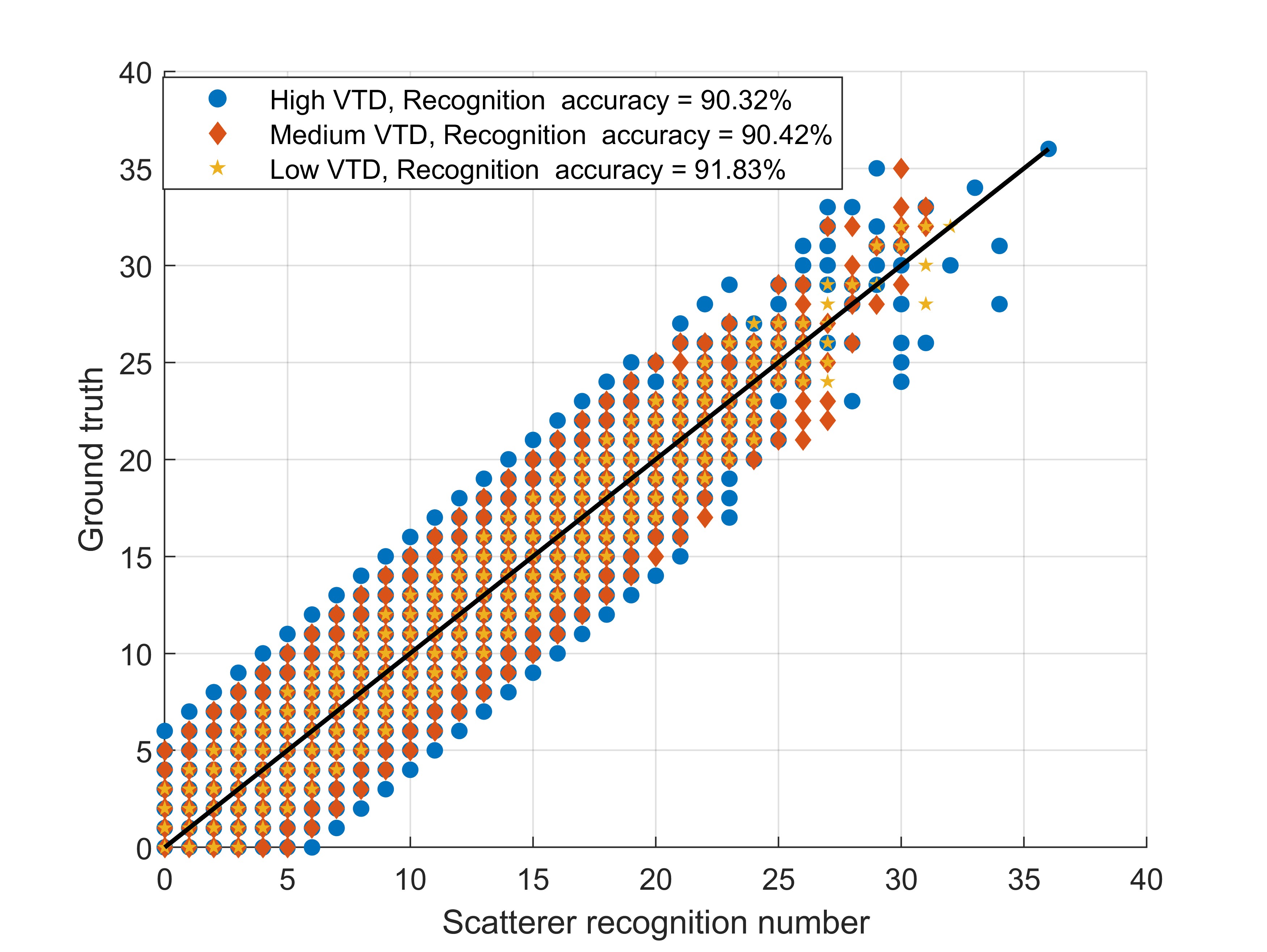}
	\caption{Comparison of scatterer recognition number and ground truth.}
	\label{number}
\end{figure}

\begin{figure}[!t]
	\centering
\includegraphics[width=0.49\textwidth]{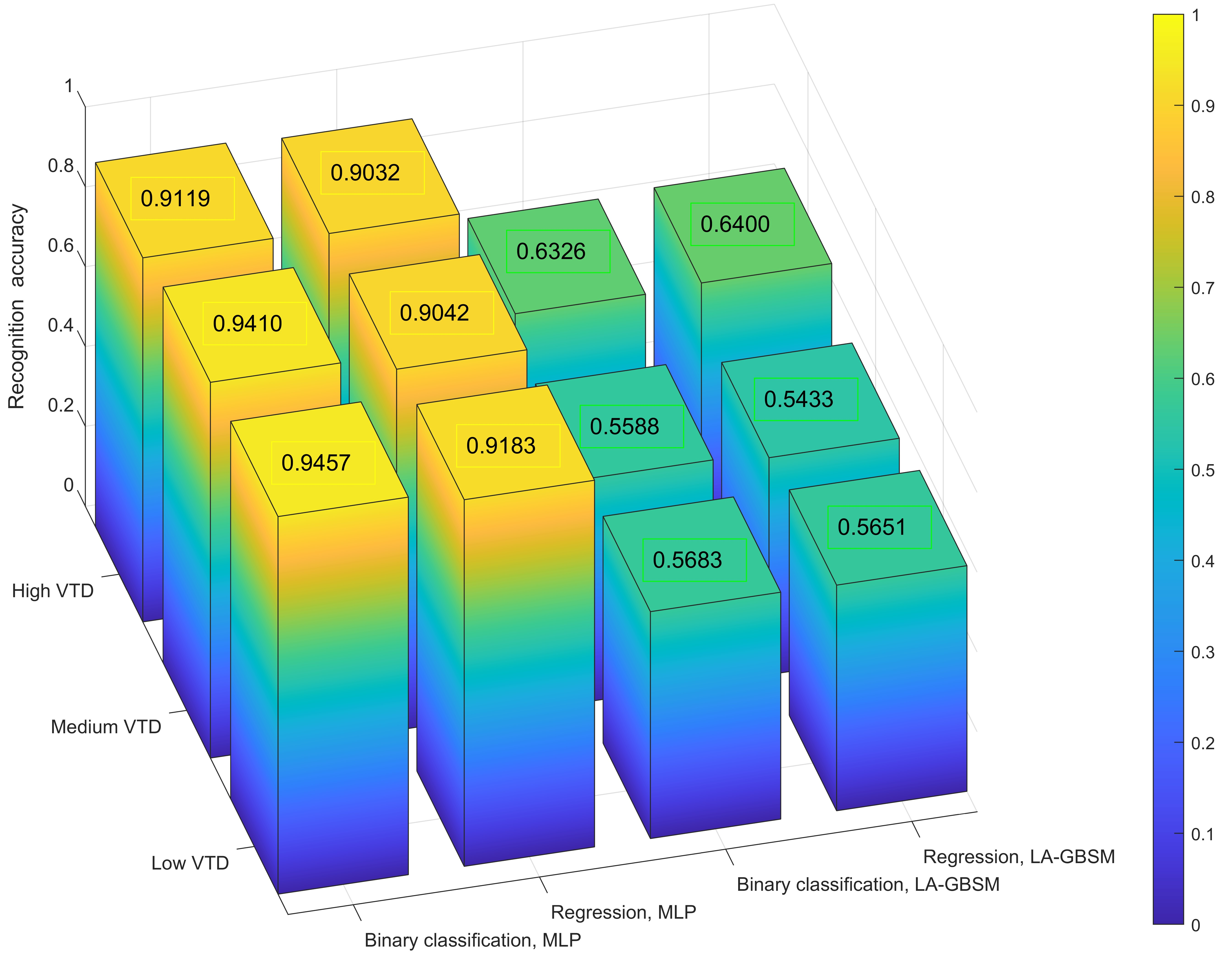}
	\caption{Comparison of the proposed approach and the random generation approach in \cite{LA-GBSM}.}
	\label{Compare LA-GBSM}
\end{figure}

\begin{figure*}[!t]
	\centering
\includegraphics[width=0.9\textwidth]{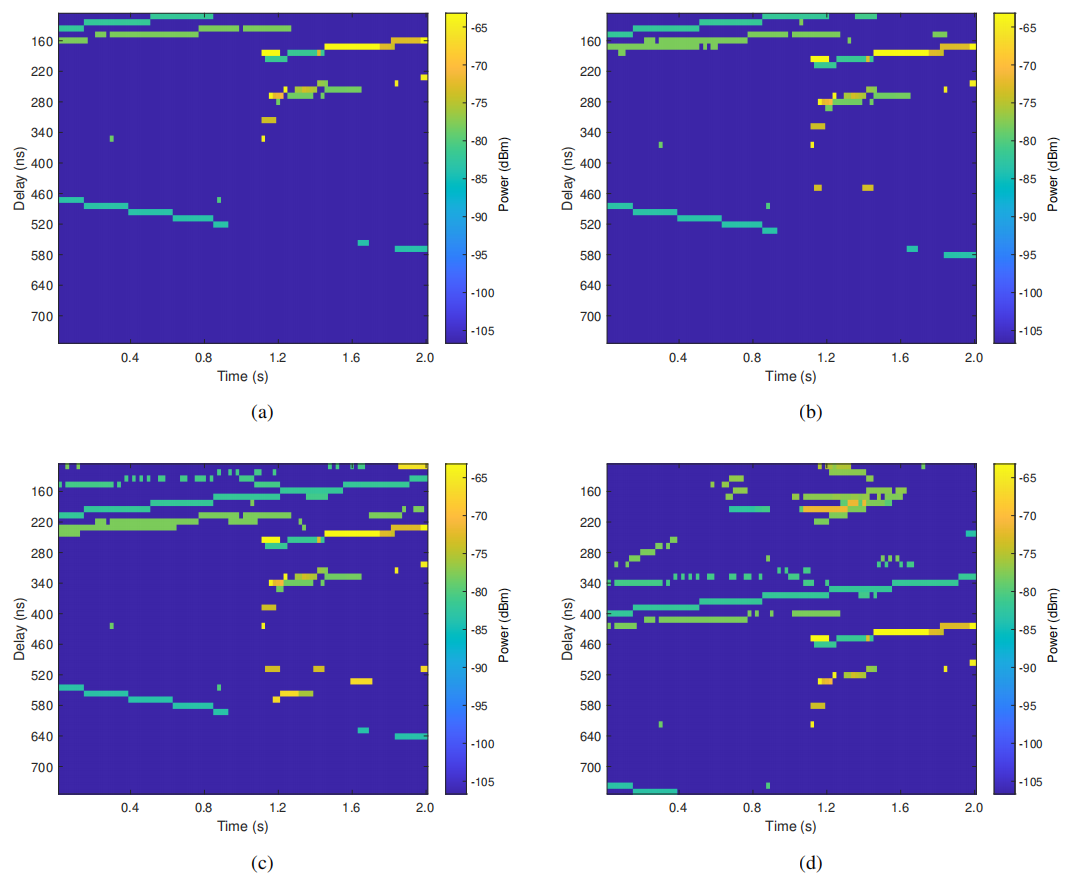}
	\caption{Comparison of PDPs. (a) RT-based results. (b) The proposed multi-modal intelligent channel model with the mapping relationship exploration. (c) The LA-GBSM without the mapping relationship exploration in \cite{LA-GBSM}. (d) The uni-modal standardized channel model in \cite{CCC}. }
	\label{PDP}
\end{figure*}

	\begin{table}[!t]
	\setlength{\abovecaptionskip}{0.1cm} 
	\renewcommand\arraystretch{0.05} 
	\centering
	\caption{Hyper-parameter setting.}
	\label{setup2}
	\begin{tabular}{c|c}
		\toprule[0.35mm]
		\textbf{Parameter}	&\textbf{Value}  \\
		\midrule[0.15mm]
		Batch size	& 16 \\ 
		\midrule[0.15mm]
		Starting learning rate & $1 \times 10^{-3}$ \\ 
		\midrule[0.15mm]			
		Learning rate scheduler	& Every 4 epochs \\ 
		\midrule[0.15mm]			
		Learning-rate decaying factor & 0.9 \\
		\midrule[0.15mm]		
		Epochs  & 200 \\ 	
		\midrule[0.15mm]		
		Optimizer  & ADAM \\ 
		\midrule[0.15mm]			
		Loss function  & MSEloss \\ 	
		\bottomrule[0.35mm]		
	\end{tabular}	
\end{table}

In Fig.~\ref{VTD-street}, the scatterer recognition accuracy in each clustering group of LiDAR point clouds with different VTDs and streets. The scatterer recognition accuracy is computed by $P = 1-\frac{N_\mathrm{error}}{N_\mathrm{all}}$, where $N_\mathrm{error}$ is the sum of differences between the recognized scatterer number and the ground truth, and $N_\mathrm{all}$ is the sum of the ground truth.
In Fig.~\ref{VTD-street}, the scatterer recognition accuracy in the aforementioned nine conditions exceeds 90\%, with an average value of 90.87\%.

Fig.~\ref{Heatmap} illustrates the probability heat map of scatterer recognition number error in the same nine conditions as Fig.~\ref{VTD-street}. From Fig.~\ref{Heatmap}, it can be seen that the cases where the scatterer recognition number differs from the ground truth by either 0 or 1 accounting for approximately 90\% of the instances.

Fig.~\ref{number} compares the scatterer recognition number and ground truth with different VTDs. Although there are many scatterers in the clustering group, the scatterer recognition accuracy exceeds 90\%. Due to the highest number of scatterers in high VTD, the scatterer recognition accuracy is the lowest.

Fig.~\ref{Compare LA-GBSM} compares the scatterer recognition accuracy of the proposed approach and the random generation approach in \cite{LA-GBSM} with different VTDs. 
In the random generation approach, the scatterer number in each clustering group is randomly generated according to the derived number distribution in \cite{LA-GBSM}. Binary classification is calculated by the recognition accuracy of whether there are scatterers on the clustering group. The regression is calculated by the recognition accuracy of the scatterer number on the clustering group. The proposed approach achieves an accuracy improvement of over 29.13\% compared to the accuracy of the random generation approach.

As the power delay profile (PDP) represents the power of received multipath components, which can be described  by scatterers, with propagation delays, Fig.~\ref{PDP} compares PDPs of RT-based results, the proposed multi-modal intelligent channel model with the mapping relationship exploration, the LA-GBSM without the mapping relationship exploration in \cite{LA-GBSM}, and the uni-modal standardized channel model in \cite{CCC}. 
The transceiver link is  Car1 (Tx) and Car8 (Rx) at the crossing street with high VTD. For a fair comparison, key parameters, such as carrier frequency, bandwidth, antenna number, and vehicular velocity, remain the same. Owing to the accurate scatterer recognition from LiDAR point clouds and the exploration of mapping relationship, the simulated PDP based on the proposed multi-modal intelligent channel model closely matches with the RT-based PDP. The close match between the simulated and RT-based PDPs together with the smooth evolution of PDPs over time demonstrate the high-precision modeling of environment-channel non-stationarity and consistency in the proposed model. However, the simulated PDPs of ignoring mapping relationship exploration \cite{LA-GBSM} and utilizing uni-modal information \cite{CCC} cannot match RT-based results.

\section{Conclusions}
This paper has proposed a novel multi-modal intelligent vehicular channel model via SoM to support the design of ITSs. In the proposed model, the SoM mechanism, i.e., mapping relationship between physical environment and electromagnetic space, has been explored based on a new dataset.  With the help of multi-modal information from communication and LiDAR devices, by leveraging  LiDAR point clouds for scatterer recognition, environment-channel non-stationarity and consistency have been modeled. Simulation results have demonstrated that the proposed approach has achieved a scatterer recognition accuracy of over 90\% and has exhibited an improvement of over 29.13\% compared to the random generation approaches. By further capturing environment-channel non-stationarity and consistency, the accuracy of the proposed multi-modal intelligent vehicular channel model has been validated by comparing RT-based and simulation results.

\ifCLASSOPTIONcaptionsoff
  \newpage
\fi

\end{document}